\documentclass[11pt]{article}

\usepackage{SIGACTNews}
\usepackage{url}

\newcommand{\COLUMNsplit}{\vskip 2em\hrule \vskip 3em}

\newcommand{\COLUMNguest}[2]{%
  \begin{center}%
   {\LARGE\bf #1 \par}%
    \vskip 1.5em%
    {\large
      \lineskip .75em%
      \begin{tabular}[t]{c}%
        #2
      \end{tabular}\par}%
    \vskip 1.5em%
  \end{center}%
  \par}

\usepackage{graphics,graphicx}
\usepackage{epsf}
\usepackage{amsmath}
\usepackage[usenames]{color}

\definecolor{bluu}{rgb}{0.15,0.19,0.54}

\def\obdds{OBDDs}
\def\obdd{OBDD}
\def\tCNF{{\it toCNF}} 
\def\SAT{{\rm SAT}} 

\def\ECTLKy {{{\textsc{ECTLK$_y$}}}}
\def\Cl{{\mathcal{X}}}  %
\def\RR{{\rm I\!R}}

\def\Pk {P_k}   

\def\ATL{${\rm ATL}$}

\def\LTL{${\rm LTL}$}

\def\CTL{${\rm CTL}$}

\def\CTLPK {{${\rm CTL_{p}K}$}}

\def\CTLK{${\rm CTLK}$}

\def\ECTLK{\rm{ECTLK}}

\def\ECTLKD {{${\rm ECTLKD}$}}

\def\PV{PV}             %

\def\Pk {{P_k}}

\def\U {{\rm U}}   %
\def\G {{\rm G}}  %
\def\X {{\rm X}}  %
\def\F {{\rm F}}  %
\def\A {{ \rm A}}
\def\E {{ \rm E}}
\def\Y {{ \rm Y}}

\def\O {{ \rm O}}

\def\H {{ \rm H}}

\def\BMC {{\rm{BMC}}}
\def\Log{{\cal L}}

\def\verics{{\rm VerICS}}

\def\QBF{{\rm QBF}}
\def\sat{\models}

\let\none = -

\def\Pk {P_k}

\def\BMC {{\rm{BMC}}}
\def\UMC {{\rm{UMC}}}

\def\E {{\rm{E}}}

\def\M {{\rm{M}}}

\def\NK {\overline{\K}}
\def\ND {\overline{\D}}
\def\NC {\overline{C}}
\def\NE {\overline{\E}}

\def\NN{{\rm I\!N}}
\def\Nd {{\rm I\!N_{+}}}

\def\PV {{\mathcal{PV}}}

\def\V {{\mathcal{V}}}

\def\U {{\rm U}} %
\def\G {{\rm G}} %
\def\X {{\rm X}} %
\def\F {{\rm F}} %

\def\A {{ \rm A}}
\def\E {{ \rm E}}
\def\Y {{ \rm Y}}

\def\O {{ \rm O}}
\def\H {{ \rm H}}

\def\CNF {{${\rm CNF}$}}

\def\LTL {{${\rm LTL}$}}
\def\CTL {{${\rm CTL}$}}

\def\CTLPK {{${\rm CTL_{p}K}$}}

\def\CTLK{{\rm{CTLK}}}
\def\CE {{\rm{CTLK}}}

\def\true{\mbox{{\bf true}}}
\def\false{\mbox{{\bf false}}}

\def\verics{Ver{\scriptsize ICS}}
\def\Cl{{\cal X}}               %
\def\BMC {{\rm{BMC}}}
\def\UMC {{\rm{UMC}}}
\def\QBF {{\rm{QBF}}}
\def\CNF {{\rm{CNF}}}

\def\Pk {P_k}               %

\def\true{\mbox{{\bf true}}}
\def\false{\mbox{{\bf false}}}

\def\verics{Ver{\scriptsize ICS}}   %

\def\iota{g^0}

\def\mas{multi-agent system}

\def\CE {{\CTLPK}}

\def\BMC {{\rm{BMC}}}
\def\UMC {{\rm{UMC}}}

\def\E {{\rm{E}}}

\def\M {{\rm{M}}}
\def\NK {\overline{K}}
\def\ND {\overline{D}}
\def\NC {\overline{C}}
\def\NE {\overline{E}}

\def\NN{{\rm I\!N}}
\def\Nd {{\rm I\!N_{+}}}

\def\PV {{\mathcal{PV}}}

\def\V {{\mathcal{V}}}

\def\MDC#1 {\M_{DC_{#1}}}

\def\MCDC#1#2 {\M_{CDC_{#1}}^{#2}}

\let\sat=\models
\let\phi=\varphi

\let\implies=\rightarrow

\let\none = -

\def\BMC{BMC}
\def\UMC{UMC}
\def\QBF{QBF}

\def\PV{PV}                             %

\def\CTL{${\rm CTL}$}

\def\LTL{${\rm LTL}$}

\def\CTLK{${\rm CTLK}$}
\def\ECTLK{${\rm ECTLK}$}

\def\CTLpK{${\rm CTL_{p}K}$}

\def\ECTLKD{${\rm ECTLKD}$}

\def\TECTLK{${\rm TECTLK}$}

\def\X{{\rm X}}                 
\def\A{{\rm A}}
\def\E{{\rm E}}
\def\F{{\rm F}}
\def\G{{\rm G}}
\def\U{{\rm U}}

\title{SIGACT News Logic Column 19}
\author{Riccardo Pucella\\
Northeastern University\\
Boston, MA 02115 USA\\
riccardo@ccs.neu.edu}
\date{}

\begin{document}

\SIGACTmaketitle

In this issue, Lomuscio and Penczek survey some of the recent work in
  verification of temporal-epistemic logic via symbolic model
  checking, focussing on OBDD-based and SAT-based approaches for
  epistemic logics built on discrete and real-time branching time
  temporal logics. 

On this topic, I should mention the following paper, which compares
several model checkers for epistemic logics with a temporal component,
using as a test case the Russian Cards problem:
\begin{quote}
H.~P. van Ditmarsch, W.~van~der Hoek, R.~van~der Meyden, and J.~Ruan.
Model checking Russian Cards.
{\em Electronic Notes in Theoretical Computer Science},
  149(2):105--123, 2006.
\end{quote}
The Russian Card problem is described here:
\begin{quote}
H.~P. van Ditmarsch.
The {Russian} {Cards} problem.
{\em Studia Logica}, 75:31--62, 2003.
\end{quote}

\COLUMNsplit

\COLUMNguest{Symbolic Model Checking for Temporal-Epistemic Logics\footnote{\copyright{} A. Lomuscio and W. Penczek, 2007.}}
{\begin{tabular}{ccc}
Alessio Lomuscio  &  \quad &             Wojciech Penczek\\
Department of Computing &&    Institute of Computer Science\\
Imperial College London&&    Polish Academy of Sciences\\
London, UK  &&   Warsaw, Poland\\
 && and \\
&& Podlasie Academy\\
&& Siedlce, Poland\end{tabular}}

\section{Introduction}

The study of epistemic logics, or logics for the representation of
knowledge, has a long and successful tradition in Logic, Computer
Science, Economics and Philosophy. Its main motivational thrust is the
observation that knowledge of the principals (or \emph{agents}) in an
exchange is fundamental in the study not only of the information they
have at their disposal, but also in the analysis of their rational
actions and, consequently, of the overall behaviour of the system.  It
is often remarked that the first systematic attempts to develop modal
formalisms for knowledge date back to the sixties and seventies and in
particular to the works of Hintikka \cite{jh:kb:cp} and Gettier
\cite{Lenzen78}.  The line of work at the time focussed on the adequacy of
particular principles, expressed as axioms of modal logic,
representing certain properties of knowledge in a rational setting.
The standard framework consisted of the propositional normal modal
logic $S5_n$ \cite{blackburn+01} built on top of the propositional
calculus by considering the axioms 
\begin{align*}
 K: \quad & K_i (p \implies q) \implies K_i p \implies K_i q\\
 T: \quad & K_i p \implies p\\
 4: \quad & K_i p \implies K_i K_i p\\
 5: \quad & K_i p \implies K_i \lnot K_i p,
\end{align*}
together with usual normal
rules of  necessitation $Nec: \mbox{From } \phi \mbox{ infer } K_i
\phi$ and modus ponens.  Since then several other formalisms have been
introduced accounting for weaker notions of knowledge as well as
subtly different mental notions such as belief, explicit knowledge and
others.

While in the sixties soundness and completeness of these formalisms
were shown, the standard semantics considered was the one of plain
Kripke models.  These are models of the form $M = (W,\{R_i\}_{i \in
  A}, V)$, where $W$ is a set of ``possible worlds'', $R_i \subseteq W
\times W$ is a binary relation between worlds expressing epistemic
indistinguishably between them, and $V: W \to 2^{\PV}$ is an
interpretation function for a set of basic propositional variables
$\PV$.  Indeed, much of the theory of modal logic has been developed
in this setting up to recent times.  However, in the eighties and
nineties attention was given to finer grained semantics that accounted
for the particular states of computation in a system.  In terms of
epistemic logic the challenge was to develop semantics that accounted
both to the low-level models of (a-)synchronous actions and protocols,
and that at the same time would be amenable to simple yet intuitive
notions of knowledge.  The key basic semantical concept put forward at
the time satisfying these considerations was the one which became
popular with the name of \emph{interpreted system}.  Originally developed
independently by Parikh and Ramanujam \cite{ParikhRamanujam85},
Halpern and Moses \cite{HalpernMoses90} and Rosenscheim
\cite{Rosenschein85} and later popularised by \cite{fhmv:rak}, the
interpreted system model offered a natural yet powerful formalism to
represent the temporal evolution of a system as well as the evolution
of knowledge of the principals in the run.  The development of this model
triggered a tremendous acceleration in the study of logics for
knowledge with several results being produced both in terms of
axiomatisations with respect to several different classes of models of
agents (synchronous, asynchronous, perfect recall, no learning, etc.)
as well as applications of these to standard problems such as
coordinated attack, communication, security, and others.

  In this setting logic was most often seen as a formal reasoning
  tool.  Attention was given to the exploration of metaproperties of
  the various formalisms (such as their completeness, decidability, and
  computational complexity), axiomatisations developed.  Attempts were
  made to verify systems automatically by exploring the relation
  $\Gamma \vdash_L \phi$, where $\phi$ is a specification for the
  system, $L$ is the axiomatised logic representing the system and
  $\Gamma$, a set of formulae expressing the initial conditions.
  However, partly due to the inherent complexity of some of the
  epistemic formalisms, verification of concrete systems via theorem
  proving for epistemic logics did not attract too much attention.

  At the same time (the early nineties) the area of verification by
  model checking \cite{CGP99} began acquiring considerable attention
  with a stream of results being produced for a variety of temporal
  logics.  The idea of switching attention from theorem proving to
  model checking became prominent \cite{halpernvardi:manifesto}.
  However, it was not before the very end of the nineties that similar
  ideas began becoming applied to the verification of multi-agent
  systems via temporal-epistemic formalisms.  The first contribution
  in the area to our knowledge dates back to a paper by van der
  Meyden and Shilov \cite{meydenshilov99}, where the complexity of
  model checking perfect recall semantics is analysed.  After that
  attention switched to the possible use of ad-hoc local propositions
  for translating the verification of temporal-epistemic into plain
  temporal logic \cite{HoekWooldridge02a}.  Following this there were
  studies on the extension of bounded model checking algorithms
  \cite{PenczekLomuscio03b} and binary-decision diagrams
  \cite{RaimondiLomuscio05b}.  Several other extensions and algorithms
  later appeared.

  The aim of this paper is to survey some of the results by the
  authors in this area over the past few years.  The area has grown
  tremendously and it is impossible to provide a comprehensive yet
  technical enough survey in a relatively compact article; some other
  approaches are discussed in Section~\ref{relatedwork}, but others,
  inevitably, are unfortunately left out.  In particular here we only
  consider approaches where knowledge is treated as a full-fledged
  modality interpreted on sets of global states in possible executions
  and not as a simple predicate as other approaches have featured.
  Concretely, the rest of the paper is organised as follows.  In
  Section \ref{ss} we present syntax and semantics of the basic logic.
  In Section \ref{bdd} we introduce and discuss an OBDD-based approach
  to verification of temporal-epistemic logic.  In Section \ref{sat}
  an alternative yet complementary approach based on bounded and
  unbounded model checking is discussed.  In Section \ref{tctlk}
  extensions to real-time are summarised briefly.  Related work is
  discussed in Section \ref{relatedwork}.

\section{Syntax and Semantics}
\label{ss}

Many model checking approaches differ depending on the syntax
supported as a specification language for the properties to be
verified by the model checker. We begin here with the basic temporal
branching time temporal-epistemic logic.

\subsection{Syntax}

Given a set of agents $A=\{1, \dots, n\}$ and a set of propositional 
variables $\PV$, we define the language $\Log$ of \CTLK\ as the fusion
between the branching time logic \CTL\ and the epistemic logic $S5_n$
for $n$ modalities of knowledge $K_i$ ($i=1,\dots,n$) and group
epistemic modalities $E_\Gamma$, $D_\Gamma$, and $C_\Gamma$ ($\Gamma \subseteq A$):
$$
\phi,\psi::= p\in \PV\ \mid \lnot \phi \mid \phi \land \psi \mid K_i
\phi \mid E_{\Gamma} \phi \mid D_{\Gamma} \phi \mid C_{\Gamma} \phi
\mid \A\X \phi \mid \A\G \phi \mid \A(\phi \U \psi) $$
In addition to the standard Boolean connectives the syntax above
defines two fragments: an epistemic and a temporal one.  The epistemic
part includes formulas of the form $K_i \phi$ representing ``agent $i$
knows that $\phi$'', $E_{\Gamma} \phi$ standing for ``everyone in
group $\Gamma$ knows that $\phi$'', $D_{\Gamma} \phi$ representing ``it
is distributed knowledge in group $\Gamma$ that $\phi$ is true'',
$C_{\Gamma}$ formalising ``it is common knowledge in group $\Gamma$
that $\phi$''. We refer to \cite{fhmv:rak} for a discussion of these
concepts and examples. 
The temporal fragment defines formulas of the
form $\A\X \phi$ meaning ``in all possible paths at each possible next 
step $\phi$ holds true''; 
$\A\G \phi$ standing for ``in all possible 
paths along $\phi$ is always true''; and $\A (\phi \U \psi)$
representing ``in all possible paths at some point $\psi$ holds true
and before then $\phi$ is true along the path''.

Whenever $\Gamma=A$ we will omit the subscript from the group
modalities $E$, $D$, and $C$. 
As customary we will also use ``diamond modalities'', i.e., 
modalities dual to the ones defined.  
In particular, for the temporal part we use $\E\F \phi = \lnot \A\G \lnot
\phi$, $\E\X \phi = \lnot \A\X \lnot \phi$ representing ``there exists
a path where at some point $\phi$ is true'' and ``there exists a path
in which at the next step $\phi$ is true'' respectively.  We will also
use the $\E(\phi \U \psi)$ with obvious meaning.  For the epistemic
part we use overlines to indicate the epistemic diamonds; in
particular we use $\NK_i \phi$ as a shortcut for $\lnot K_i \lnot
\phi$, meaning ``agent $i$ considers it possible that $\phi$'' and
similarly for $\NE_\Gamma$, $\ND_\Gamma$, and $\NC_\Gamma$.

Formulas including both temporal and epistemic modalities can
represent expressive specifications in particular scenarios, e.g., the
evolution of private and group knowledge over time, knowledge about a
changing environment as well as knowledge about other agents'
knowledge.  We refer to \cite{fhmv:rak} for standard examples such as
alternating bit protocol, attacking generals, message passing systems, etc.

\subsection{Interpreted systems semantics}

In what follows the syntax of the specification language supported is
interpreted on the multi-agent semantics of interpreted systems
\cite{fhmv:rak}. Interpreted systems are a fine-grained semantics put
forward in \cite{HalpernMoses90} to represent temporal evolution and
knowledge in multi-agent systems. Although initially developed for
linear time, given the applications of this paper we present them in
their branching time version. 
Given the model checking algorithms described later we summarise 
the formalism below in relation to a branching time model. 
For more details we refer to \cite{fhmv:rak}.

Assume a set of \emph{possible local states} $L_i$ for each agent $i$
in a set $A=\{1, \dots, n\}$ and a set $L_e$ of possible local states
for the environment $e$.  The set of \emph{possible global states} $G
\subseteq L_1 \times \dots \times L_n \times L_e$ is the set of all
possible tuples $(l_1, \dots, l_n, l_e)$ representing a snapshot of
the system as a whole.  The model stipulates that each agent $i$
performs one of the enabled actions in a given state according to a
\emph{protocol function} $P_i: L_i \to 2^{Act_i}$.  $P_i$ maps local
states to sets of possible actions for agent $i$ within a repertoire
of its actions $Act_i$.  Similarly, the environment $e$ is assumed to
be performing actions following its protocol $P_e: L_e \to 2^{Act_e}$.
\emph{Joint actions} $(act_1, \dots, act_n, act_e)$ are tuples of
actions performed jointly by all agents and the environment in
accordance with their respective protocol.  Joint actions are used
to determine the transition function $T \subseteq G \times Act_1
\times \dots \times Act_n \times Act_e \times G$ which gives the
evolution of a system from an initial global state $g^0 \in G$.  A
{\em path} \mbox{$\pi = (g_{0}, g_{1}, \ldots)$} is a maximal sequence
of global states such that \mbox{$(g_{k}, g_{k+1}) \in T$} for each $k
\geq 0$ (if $\pi$ is finite then the range of $k$ is restricted accordingly).
For a path $\pi=(g_{0}, g_{1}, \ldots)$, we take $\pi(k) = g_{k}$.
By $\Pi(g)$ we denote the set of all the paths starting at $g \in G$.

The model above can be enriched in several ways by 
expressing explicitly observation functions for the agents in the
system or by taking more concrete definitions of the sets of local
states thereby modelling specific classes of systems (perfect recall,
no learning, etc.). We do not discuss these options here; we simply
note that in a later section we will pair this semantics with an
automata-based one.

To interpret the formulas of the language $\Log$ for convenience we
define models simply as tuples $M=(G,g^0,T,\sim_1,\dots,\sim_n,V)$,
where $G$ is the set of the global states reachable from the initial 
global state $g^0$ via $T$;
$\sim_i\; \subseteq G \times G$ is an epistemic relation for agent $i$ 
defined by $g \sim_i g'$ iff $l_i(g)=l_i(g')$, 
where $l_i: G \to L_i$ returns the local state of agent $i$ given a global state; 
and $V: G \times \PV\ \to \{true, false\}$ is an interpretation 
for the propositional variables $\PV$ in the language.

The intuition behind the definition of models above is that the global 
states whose local components are the same for agent $i$ are not
distinguishable for the agent in question. 
This definition is standard in epistemic logics via interpreted
systems---again we refer to \cite{fhmv:rak} for more details.  

We can use the model above to give a satisfaction relation $\sat$
for $\Log$ inductively as standard. 
Let $M$ be a model, $g =(l_1,\ldots,l_n)$ a global state, and $\phi, \psi$ 
formulas in $\Log$: 
\begin{itemize}
\item $(M,g) \sat p$ iff $V(g,p)=true$, 
\item $(M,g) \sat K_i \phi$ iff for all $g' \in G$ 
      if $g \sim_i g'$, then $(M,g') \sat \phi$,
\item $(M,g) \sat D_{\Gamma} \phi$ iff for all $i \in \Gamma$ and
      $g' \in G$ if $g \sim_i g'$, then $(M,g') \sat \phi$,
\item $(M,g) \sat E_{\Gamma} \phi$ iff $(M,g) \sat \bigwedge_{i \in
    \Gamma} K_i \phi$,
\item $(M,g) \sat C_{\Gamma} \phi$ iff for all $k\geq 0$ we have
  $(M,g) \sat E_{\Gamma}^k \phi$,
\item $(M,g) \sat \A\X \phi$ iff for all $\pi \in \Pi(g)$ we have
      $(M,\pi(1)) \sat \phi$,
\item $(M,g) \sat \A\G \phi$ iff for all $\pi \in \Pi(g)$ and for all
      $k \geq 0$ we have $(M,\pi(k)) \sat \phi$,
\item $(M,g) \sat \A(\phi \U \psi)$ iff for all $\pi \in \Pi(g))$
      there exists a $k \geq 0$ such that  $(M,\pi(k)) \sat \psi$ and for
      all $0\leq j < k$ we have $(M,\pi(j)) \sat \phi$.
\end{itemize}
The definitions for the Boolean connectives and the other inherited
modalities are given as standard and not repeated here. 
$E^k \phi$ is to be understood as a shortcut for $k$ occurrences of 
the $E$ modality followed by $\phi$, i.e., 
$E^0 \phi = \phi$; 
$E^1 \phi = E \phi$; 
$E^{k+1} \phi =E E^k \phi$.
  
\subsection{The dining cryptographers problem}

The formalism of interpreted systems has been used successfully to
model a variety of scenarios ranging from basic communication
protocols (e.g., the bit transmission problem, message passing
systems), to coordination (e.g., the attacking generals setting),
deadlocks (e.g., the train-gate-controller scenario), etc. 
We refer the reader to the specialised literature; the key consideration here
is that in each of these scenarios it is shown that temporal-epistemic
languages can be used to express specification for the systems and the
individual agents very naturally.

To exemplify this we present a protocol for anonymous broadcast very
well-known in the security literature: the dining cryptographers (DC).
The DC was introduced by Chaum~\cite{Chaum} and analysed in a
temporal-epistemic setting by Meyden and Su \cite{meydensu02}.  A
reformulation to include cheating cryptographers (see
Section~\ref{relatedwork}) appears in \cite{Kacprzak+06}.  We report
the original wording here \cite{Chaum} (part of this text was
originally cited in \cite{meydensu02}).

\begin{quote}
{\it Three cryptographers are sitting down to dinner at their
favorite three-star restaurant. Their waiter informs them that
arrangements have been made with the maitre d'hotel for the bill
to be paid anonymously. One of the cryptographers might be paying
for dinner, or it might have been NSA (U.S. National Security
Agency). The three cryptographers respect each other's right to
make an anonymous payment, but they wonder if NSA is paying. They
resolve their uncertainty fairly by carrying out the following
protocol:

Each cryptographer flips an unbiased coin behind his menu, between
him and the cryptographer on his right, so that only the two of
them can see the outcome. Each cryptographer then states aloud
whether the two coins he can see--the one he flipped and the one
his left-hand neighbor flipped--fell on the same side or on
different sides. If one of the cryptographers is the payer, he
states the opposite of what he sees. An odd number of differences
uttered at the table indicates that a cryptographer is paying; an
even number indicates that NSA is paying (assuming that dinner was
paid for only once). Yet if a cryptographer is paying, neither of
the other two learns anything from the utterances about which
cryptographer it is.}
\end{quote}

Temporal-epistemic logic can be used to analyse the specification of
the example---we summarise here the description reported in
\cite{RaimondiLomuscio05b,Raimondi06}.  It is relatively
straightforward to model the protocol above by means of interpreted
systems. For each agent $i$ we can consider a local state consisting
of the triple $(l_i^1,l_i^2,l_i^3)$, representing respectively whether
the coins observed are the same or different, whether agent $i$ paid
for the bill, and whether the announcements have an even or odd
parity.  A local state for the environment can be taken as a 4-tuple
$(l_e^1,l_e^2, l_e^3, l_e^4)$ where $l_e^1$, $l_e^2$, $l_e^3$ represent the coin
tosses for each agent and $l_e^4$ represents whether or not the agent in
question paid for the bill. Actions and protocols for the agents and
the environment can easily be given following Chaum's narrative
description above and relations for the temporal evolution and the
epistemic relation easily built in this way.

In principle by coding the above we would be able to show on the model
for DC that 
\[
(M_{DC}, g^0) \sat \bigwedge_{i\in A} ({\bf odd} \land \lnot {\bf paid_i}) 
\implies 
\A\X(K_i (\bigvee_{j\neq i} {\bf paid_j}) \bigwedge_{k \neq i} \lnot K_i {\bf paid_k})
\]
The specification above states that if an agent $i$ observes an odd
parity and did not cover the bill then in all next states (i.e., when
the announcements have been made) she will know that one of the others
paid for dinner but without knowing who it was.

Although conceptually easy, the example is already large enough to
make it difficult to work out all possible execution traces on the
model.  Of note is the fact that DC can actually be scaled to any
number of cryptographers. By using model checking techniques 
one can verify DC up to 8 and more cryptographers with resulting state spaces
for the model of about $10^{36}$ states, and considerably more
cryptographers if the representation of the model is optimised
\cite{Kacprzak+06}. 

Other examples are equally amenable to representation via interpreted
systems and model checking via the techniques presented below.

\section{OBDD-based symbolic model checking} 
\label{bdd}

As it is customary in model checking in the following we analyse 
systems of finite states only.
Given a system $S$ and a property $P$ to be checked, the model checking
approach suggests coding $S$ as a logical model $M_S$, the property
$P$ as a logic formula $\phi_P$, and investigating whether $M_S \sat
\phi_P$. In the traditional approach the model $M_S$ is finite and
represents all the possible computations of system $S$ and $\phi_P$
is a formula in temporal logic expressing some property to be checked
on the system, e.g., liveness, safety, etc. 
When $\phi_P$ is given in \LTL\ or \CTL\ checking $\phi_P$ on an 
explicitly given $M_S$ is, of course, a very tractable problem.  
However it is impractical to represent $M_S$ explicitly,  
so $M_S$ is normally implicitly given by means of a dedicated programming 
language using imperative commands on sets of variables. 
This can be convenient for the programmer, but the number of states in the 
resulting model grows exponentially with the number of variables used 
in the program describing $M_S$ potentially causing great difficulty 
(\emph{state explosion problem}).

Much of the model checking literature in plain temporal logic deals
with techniques to limit the impact of this, the most prominent being
partial order reductions \cite{Peled93,GKPP99}, 
symmetry reductions \cite{CFJ93,EJ93,emerson},
ordered-binary decision diagrams \cite{bcmdh90,McMillan93}, 
bounded and unbounded model checking \cite{symbolic-without-bdd,McMillan02},
and (predicate) abstraction \cite{DGD94,Ball+01}.
By using partial-order reduction techniques the computational tree $M_S$ 
is pruned and certain provably redundant states eliminated and/or collapsed with others depending on the formula to be checked thereby reducing the state space.
Symmetry reductions are used to reducing the state spaces of distributed 
systems composed of many similar processes.
Predicate abstraction is based on the identification of certain
predicates which have no impact on the verification of the formula in
question; crucially it is used in verification of infinite-state
systems.  Binary-decision diagrams (described below) offer a compact
representation for Boolean formulas and traditionally constitute 
one of the leading symbolic approaches.  Bounded and unbounded model
checking (described in Subsections~\ref{bmc} and~\ref{umc}
respectively) exploit recent advances in the efficiency of checking
satisfiability for appropriate Boolean formulas suitably constructed.
Several tools have been developed for model checking temporal logic,
including SPIN \cite{Holzmann97} for partial-order reductions for
\LTL, SMV and NuSMV \cite{McMillan93,cimatti99a} for binary-decision
diagrams and bounded model checking for \LTL, and SLAM \cite{slam} for
partial-ordered reductions for safety properties.  Several other tools
exist for other varieties of temporal logic, e.g., real-time logics,
probabilistic temporal logic, and indeed other implementations are
available for the same or slightly different techniques.

Even if all tools mentioned above are nowadays very sophisticated and
support ad-hoc input languages they are limited to temporal logics only. 
In the rest of the paper we discuss techniques and tools supporting
temporal-epistemic logics.

\subsection{The ordered binary decision diagrams approach}

The two main model checking platforms for temporal-epistemic logic
based on binary-decision diagrams are the MCK and the MCMAS toolkits.
Both in their experimental phase, they implement model checking of
temporal-epistemic logic on interpreted systems semantics via
ordered-binary decision diagrams. MCK \cite{mck04,mck-tool} implements
a variety of different semantics (observational, perfect recall, etc),
supports a concise and specialised input language, and was the first
model checker available supporting temporal-epistemic logic.  MCMAS
\cite{RaimondiLomuscio03a,mcmas-tacas06} implements standard
interpreted systems semantics and a number of extensions, including
deontic modalities, explicit knowledge, \ATL, etc. In terms of
implementations the two tools are rather different. MCK is implemented
in Haskell using Long's BDD library (written in C), whereas MCMAS is
implemented in C++ and relies on Somenzi's \cite{cudd-2.4.0} BDD
package (also in C).  MCMAS and its theoretical background is
succinctly described in the rest of this section; we refer to
\cite{Raimondi06} for an in-depth description.

Irrespective of the implementation details the angle when working on
ordered-binary decision diagrams (\obdds) is the symbolic
(OBDD-based) representation of sets and functions paired with the
observation that to work out whether $(M,g) \sat \phi$ it is
sufficient to evaluate whether or not $g \in SAT(\phi)$ where
$SAT(\phi)$ is the set of states in the model $M$ satisfying $\phi$.
To introduce the main ideas of the approach we proceed in three
stages: first, we observe we can code sets as Boolean formulas;
second, we show how \obdds\ offer a compact representation to Boolean
functions; third we give algorithms for the calculation of $SAT(\phi)$.

First of all observe that given a set $G$ of size $|G|$ it is obvious
how
to associate uniquely a vector of Boolean variables $(w_1,\ldots,w_m)$
to any element $g \in G$ where $m=\lceil log_2 |G| \rceil$. (Note that
a tuple of $m$ places can represent $2^m$ different elements).  Any
subset $S \subseteq G$ can be represented by using a characteristic
function $f_S: (g_1, \dots, g_m) \to \{0,1\}$, expressing whether the
element (as encoded) is in $S$ or not.  Note that functions and
relations can also be encoded as Boolean functions; for instance to
encode that two states are related by some relation we can simply
consider a vector of Boolean functions comprising of two copies of the
representation of the state to which we add a further Boolean variable
expressing whether or not the states are related.  Vectors designed in
this way represent conjunctions of Boolean atoms or their negation and
as such constitute a simple (albeit possibly long) Boolean formula.

\begin{figure}
\centerline{\includegraphics[width=12cm]{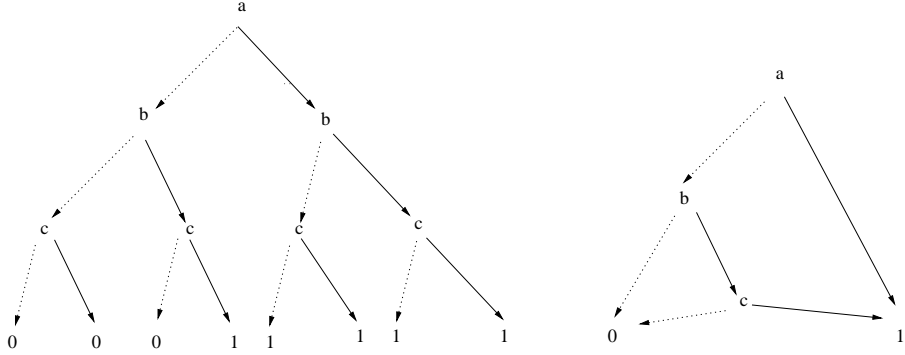}}
\caption{A BDT for the Boolean function $a \lor (b \land c)$ (left)
  and its corresponding BDD (right). The dotted lines correspond to
  assigning the value false to the atom whose name the edge leaves
  from. Conversely solid lines represent assignments to true.}
\label{bdt}
\end{figure}

In the construction of OBDD-based model checking for plain temporal
logic it is normally assumed that the propositions themselves
(appropriately ordered) constitute the basis for the encoding of the
states of the model. In the MCMAS approach Boolean functions first and
then \obdds\ are constructed iteratively by considering all aspects of
the interpreted system given. These involve building the:
\begin{itemize}
\item Boolean functions for the sets of local, global states,
  actions, initial global states;
\item Boolean functions representing the protocols for each agent, the
  local evaluation function for each agent, the valuation for the
  atoms;
\item Boolean functions representing the global temporal relation and
  the $n$ epistemic relations for the agents. The Boolean formula
  coding the temporal relation needs to encode that joint actions
  correspond to enabled actions for all agents: $f_T(g,g') =
  \bigvee_{a \in JointAct} (g,a,g') \in T \bigwedge_{i \in A} a_i \in
  P_i(l_i(g))$, where $a=(a_1, \dots, a_n)$ is a joint action for the
  system and all individual action components $a_i$ are enabled by the
  local protocols at the corresponding local state $l_i(g)$ in $g$.
  The epistemic relations for the agents can be represented simply by
  imposing equality on the corresponding local state component.

\item A Boolean formula representing the set of reachable states for
  the interpreted system. This can be encoded as standard by
  calculating the fix-point of the operator 
$\tau(Q)=(I(g) \lor \exists g'(T(g,a,g') \land Q(g')).$
\end{itemize}
Boolean functions are a convenient representation to perform certain
logical operations on them (e.g., $\land, \lor$); however it is well
known that working out their satisfiability and validity can be
expensive.  Truth tables themselves do not offer any advantage in this
respect: for instance checking satisfiability on them may involve
checking $2^n$ rows of the table where $n$ is the number of atoms
present. \obdds\ constitute a symbolic representation for Boolean
functions and are normally much cheaper to handle. Before introducing
\obdds\ observe that to every Boolean function we can associate a
binary decision tree (BDT), in which each level represents a different
atom appearing in the Boolean function. Taking a different path along
the tree corresponds to selecting a particular combination of values
for the atoms (see Figure~\ref{bdt}), thereby determining the truth
value of the formula.

In most instances a BDT is not an efficient representation of its
corresponding Boolean function. However, a series of operations can be
performed on it to reduce it to a binary decision diagram (BDD). 
A BDD is a directed acyclic graph with an initial node, and in which each
node (representing a Boolean atom) has two edges (corresponding to
decision points true and false) originating from it with the final
leaves being either ``true'' or ``false'' (see Figure~\ref{bdt}).
There are several algorithms for producing BDDs from BDTs; however the
order of the operations on the initial BDT affects the resulting BDD
and, most crucially, comparing BDDs turns out to be an expensive
operation. What makes the whole approach useful is the provable
assertion that there exist sets of algorithms computing canonical BDDs
once the ordering of the variables is fixed. In other words, as long
as the ordering of the variables is fixed the resulting BDD is unique
for a given Boolean function. This is a remarkable result and leads to
an alternative technique to compare Boolean functions: compute their
canonical BDDs; if they are the same they represent the same function,
if not they are the result of different functions. The canonical BDDs
produced by this set of algorithms are normally referred to as \obdds\
and constitute one of the leading data structures in symbolic model
checking. We do not discuss algorithms to manipulate BDDs here and
refer to \cite{HuthRyan00} for details; but of particular significance
is the fact that Boolean operations on Boolean functions can be done
directly on the corresponding \obdds\ without a very significant loss
in performance. Other model-checking specific set operations such as
computing pre-images (see below) may also be coded in terms of the
corresponding BDDs. For more details on \obdds\ and related techniques
we refer to \cite[Chapter 6]{HuthRyan00} and references, notably
\cite{bryant-bdd}.

We now present the algorithms for the calculation of the set of states
$SAT(\phi)$ satisfying a formula $\phi$ in $\Log$. In the \obdd\
approach all sets of states below are computed symbolically on the
corresponding \obdds.
 
\begin{small}
\begin{center}
\fbox{
\begin{tabular}{l}
  $SAT(\phi)$ \{ \\
\ \ $\phi$ is an atomic formula: return $\{g \mid V(g,\phi)=true\}$;\\
\ \ $\phi$ is $\neg \phi_1$: return $S \ \backslash \ SAT(\phi_1)$;\\
\ \ $\phi$ is $\phi_1 \land \phi_2$: return $SAT(\phi_1) \cap SAT(\phi_2)$;\\
\ \ $\phi$ is $\E\X \phi_1$: return $SAT_{EX}(\phi_1)$;\\
\ \ $\phi$ is $\E(\phi_1 \U \phi_2)$: return $SAT_{EU}(\phi_1,\phi_2)$;\\
\ \ $\phi$ is $\E\F\phi_1$: return $SAT_{EF}(\phi_1)$;\\
\ \ $\phi$ is $K_i \phi_1$: return $SAT_{K}(\phi_1,i)$;\\
\ \ $\phi$ is $E_\Gamma \phi_1$: return $SAT_{E}(\phi_1,\Gamma)$;\\
\ \ $\phi$ is $D_\Gamma \phi_1$: return $SAT_{D}(\phi_1,\Gamma)$;\\
\ \ $\phi$ is $C_\Gamma \phi_1$: return $SAT_{C}(\phi_1,\Gamma)$;\\
\}
\end{tabular}}
\end{center}
\end{small}

In the algorithm above, the auxiliary procedures $SAT_{EX}, SAT_{EU},
SAT_{EF}$ follow the standard algorithms used in temporal
logic.\footnote{For efficiency reasons the 
\CTL\ modalities implemented are typically $\E\X$, $\A\F$, and $\E\U$.}
For instance the set of global states satisfying $\E\X \phi$ is computed 
as follows (in what follows $G$ is the set of reachable states).
\begin{center}
\begin{small}
\fbox{
\begin{tabular}{l}
  $SAT_{EX}(\phi)$ \{ \\
\ \ X = $SAT(\phi)$;\\
\ \ Y = $\{g \in G \mid \exists g' \in X \textrm{ and } T(g,a,g')\}$ \\
\ \ return Y;\\
\}
\end{tabular}}
\end{small}
\end{center}

Note that the calculation of $\E\X$ involves working out the pre-image
of $T$.  The set of states satisfying the epistemic modalities are
defined as follow (note that below we use 
$\sim_{\Gamma}^E = \bigcup_{i \in \Gamma} \sim_i$ 
and 
$\sim_{\Gamma}^D = \bigcap_{i \in \Gamma} \sim_i$). 

\begin{center}
\begin{small}
\fbox{
\begin{tabular}{l}
  $SAT_{K}(\phi,i)$ \{ \\
\ \ X = $SAT(\neg\phi)$;\\
\ \ Y = $\{g \in S \mid \exists g' \in X \textrm{ and } \sim_i(g,g')\}$ \\
\ \ return $\neg$Y;\\
\}
\end{tabular}}
\fbox{
\begin{tabular}{l}
  $SAT_{E}(\phi,\Gamma)$ \{ \\
\ \ X = $SAT(\neg \phi)$;\\
\ \ Y = $\{g \in G \mid \,\sim_{\Gamma}^E(g,g') \textrm{ and } g'\in X\}$ \\
\ \ return $\neg$Y;\\
\}
\end{tabular}}
\fbox{
\begin{tabular}{l}
  $SAT_{D}(\phi,\Gamma)$ \{ \\
\ \ X = $SAT(\neg \phi)$;\\
\ \ Y = $\{g \in G \mid \,\sim_{\Gamma}^D(g,g') \textrm{ and } g'\in X\}$ \\
\ \ return $\neg$Y;\\
\}
\end{tabular}}
\fbox{
\begin{tabular}{l}
  $SAT_{C}(\phi,\Gamma)$ \{ \\
\ \ X = $SAT(\phi)$;\\
\ \ Y = $G$; \\
\ \ while ( X $\not=$ Y ) \{ \\
\ \ \ \ X = Y; \\
\ \ \ \ Y = $\{g \in G \mid\, \sim_{\Gamma}^E(g,g') \textrm{ and } g'\in Y \textrm{ and } g' \in SAT(\phi)\}$ \\
\ \ return Y;\\
\}
\end{tabular}}
\end{small}
\end{center}

The algorithm for $K_i \phi$ is similar in spirit to the CTL algorithm
for computing $\A\X \phi$: essentially we compute the pre-image under
the epistemic relation of the set of formulas not satisfying $\phi$
and negate the result. $E_{\Gamma} \phi$ (resp., $D_{\Gamma}
\phi$) is done similarly but on the $\sim_E^{\Gamma}$
(resp., $\sim_D^{\Gamma}$). For $C$ we need to use a fix-point
construction (fix-point constructions already appear in the algorithm
to compute the satisfiability of the until operator). In fact, note
that $C_{\Gamma} \phi = E_{\Gamma} (\phi \land C_{\Gamma} \phi)$, so
it can be computed by calculating the fix-point of
$\tau(Q)=SAT(E_{\Gamma} \phi \land Q)$ as in the table above.
All sets operations above are implemented on the
corresponding \obdds\ thereby producing the OBDD for $SAT(\phi)$. We
can then solve $(M,g^0) \sat \phi$ by answering the query $g^0 \in
SAT(\phi)$ on the corresponding OBDD.

\subsection{MCMAS}

MCMAS \cite{mcmas-tacas06,mcmas} is a GNU GPL tool that
implements the OBDD-based procedures of the previous subsection.
Input to the model checker is a program describing the evolutions of a
multi-agent system. The program is given in ISPL (Interpreted Systems
Programming Language), a language specialised for the specifications
of interpreted systems and some extensions. An ISPL program consists
of a sequence of declarations for agents in the system, valuation for
the atomic propositions, and formulas in \CTLK\ (other languages are
also supported---see extensions). An agent is given by explicitly
listing the local states it may be in, the local actions, protocols,
and the local evolution function. Note that the local evolution function 
$: L_i \times Act_1 \times \dots \times Act_n \to L_i$ 
gives a set of rules specifying the target local state when a certain
combination of actions is performed. An example of an ISPL fragment
describing a very simple agent is given in Figure~\ref{ispl}.

\begin{figure}
\begin{small}
\begin{verbatim}
Agent SampleAgent
 Lstate = {s0,s1,s2};
 Action = {a1,a2}
 Protocol:
   s0: {a1};
   s1: {a2};
   s2: {a1,a2};
 end Protocol
 Ev:
   s1 if ((AnotherAgent.Action=a7);
   s2 if Lstate=s1;
 end Ev
end Agent
\end{verbatim}
\end{small}
\caption{A fragment of ISPL code describing an agent.}
\label{ispl}
\end{figure}

Upon invocation the tool parses the input, builds the OBDD for
transition relation and the OBDD for the set of reachable states. This
is then used in the calculation of the OBDD for the sets of states
satisfying the formula to be verified. By comparing whether the
initial state belongs to this set the output is displayed. A graphical
and a web interface are available for the tool. MCMAS is presented in
detail in \cite{Raimondi06}.

\section{SAT-based symbolic model checking}
\label{sat}

\SAT-based model checking is the most recent symbolic approach for modal logic. 
It was motivated by a dramatic increase in efficiency of \SAT-solvers, i.e.,
algorithms solving the satisfiability problem for propositional formulas \cite{Zhang01}. 
The main idea of \SAT-based methods consists in translating the model checking
problem for a temporal-epistemic logic to the problem of satisfiability of 
a formula in propositional logic.
This formula is typically obtained by combining an encoding of the
model and of the temporal-epistemic property.
In principle, the approaches to \SAT-based symbolic verification
can be viewed as bounded (\BMC) or unbounded (\UMC).
\BMC\ applies to an existential fragment of a logic (here \ECTLK) 
on a part of the model, whereas \UMC\ is for an unrestricted 
logic (here \CTLK) on the whole model.

\subsection{Bounded Model Checking}
\label{bmc}

\BMC\ was originally introduced for verification of \LTL\ \cite{symbolic-without-bdd,BMC_Academic03} as an alternative
to approaches based on OBDDs.
Then, \BMC\ was defined for the existential fragment of the logic \CTL\ 
\cite{pwz02} and then extended to \ECTLK\ \cite{PenczekLomuscio03b}. 
\BMC\ is based on the observation that some properties of a system can be 
checked over a part of its model only. 
In the simplest case of reachability analysis, this approach consists 
in an iterative encoding of a finite symbolic path as 
a propositional formula. 
The satisfiability of the resulting propositional formula is then checked 
using an external SAT-solver. 
We present here the main definitions of \BMC\ for \ECTLK\ and 
later discuss extensions to more expressive logics. 
We refer the reader to the literature cited above for more details. 

To explain how the model checking problem for an \ECTLK\ formula is encoded
as a propositional formula, we first define $k$-models, bounded semantics 
over $k$-models, and then propositional encodings of $k$-paths in
the $k$-model and propositional encodings of the formulas.
In order to define a bounded semantics for \ECTLK\ we define $k$-models. 
Let $M =(G,g^0,T,\sim_1,\ldots,\sim_n,\V)$ be a model and $k \in \Nd$. 
The $k$-model for $M$ is defined as a structure 
$M_k =(G,g^0,\Pk,\sim_1,\ldots,\sim_n,\V)$,  
where $\Pk$ is the set of all the $k$-paths of $M$ over $G$,
where a $k$-path is the prefix of length $k$ of a path.

We need to identify $k$-paths that represent infinite paths 
so that satisfaction of $EG$ formulas in the bounded semantics 
implies their satisfaction on the unbounded one.
To this aim define the function \mbox{$loop: \Pk \to 2^{\NN}$} as: $
loop(\pi) \; = \; \{ l \; \mid \;0 \leq l \leq k \;and\;
(\pi(k),\; \pi(l)) \in T \}$,  which returns the set of indices
$l$ of $\pi$ for which there is a transition from $\pi(k)$ to $\pi(l)$.

Let $M_k$ be a $k$-model and $\alpha, \beta$ be \ECTLK\ formulas. 
$(M_{k}, g) \models \alpha$ denotes that $\alpha$ is true at the state $g$ of $M_k$.  
The bounded semantics is summarised as follows.
$(M_k,g) \models \E\X\alpha$ has the same meaning as for unbounded models.
$(M_k,g) \models \E \G \alpha$ states that there is a $k$-path $\pi$, 
which starts at $g$, all its states satisfy $\alpha$ and $\pi$ is a loop, 
which means that $g$ is a $T$-successor of one of the states of $\pi$.
The indexes of such states are given by $loop(\pi)$.
For the other modalities the bounded semantics is the same as
unbounded, 
insisting on reachability of the state satisfying
$\alpha$ on a path of length $k$.

Model checking over models can be reduced to model checking over $k$-models. 
The main idea of \BMC\ for \ECTLK\ is that checking $\varphi$ over $M_k$ 
is replaced by checking the satisfiability of the propositional formula $[M,\varphi]_{k} := [M^{\varphi, g^0}]_k \land [\varphi]_{M_k}$.
$[M^{\varphi, g^0}]_k$ represents (a part of) the model under 
consideration whereas $[\varphi]_{M_k}$ captures a number of constraints that must
be satisfied on $M_k$ for $\varphi$ to be satisfied.  
Checking satisfiability of an \ECTLK\ formula can be done by means of 
a SAT-solver.  Typically, we start with $k := 1$,
test satisfiability for the translation, and increase $k$ by one until
either $[M^{\varphi, g^0}]_k \land [\varphi]_{M_k}$ becomes
satisfiable, or $k$ reaches the maximal depth of $M$, which is bounded by $|G|$.
It can be shown that if $[M^{\varphi, g^0}]_k \land [\varphi]_{M_k}$
is satisfiable for some $k$, then $(M,g^0) \sat \varphi$, where
$M$ is the full unbounded model.

\subsubsection{Translation to SAT}
\label{trans}

We provide here some details of the translation. 
The states and the transitions of the system under consideration
are encoded similarly as for BDDs in Section \ref{bdd}.
Let $w=(w[1],\ldots,w[m])$ be sequence of propositions 
(called a \emph{global state variable}) for encoding global states.
A sequence $w_{0,j},\ldots,w_{k,j}$ of global state variables 
is called a symbolic $k$-path $j$.
Since a model for a branching time formula is a tree (a set of paths),
we need to use a set of symbolic $k$-paths to encode it. 
The number of them depends on the value of $k$ and the formula 
$\varphi$, and it is computed using the function $f_k$.
This function determines the number of $k$-paths  
sufficient for checking an \ECTLK\ formula, 
see \cite{wlp-lcmas04} for more details.
Intuitively, each nesting of an epistemic or temporal formula 
in $\varphi$ increases the value of $f_k(\varphi)$ by $1$, 
whereas subformulas $\E\U$, $\E\G$ and $\NC_\Gamma$ add more $k$-paths. 

The propositional formula $[M^{\varphi, g^0}]_k$, representing the
$k$-paths in the $k$-model, is defined as follows:

$$ [M^{\varphi, g^0}]_{k} := I_{g^0}(w_{0,0})\land
\bigwedge_{j=1}^{f_k(\varphi)} \bigwedge^{k-1}_{i=0}\;
T(w_{i,j},w_{i+1,j}),$$ 
where $w_{0,0}$ and $w_{i,j}$ for $0 \leq i \leq k$ 
and $1 \leq j \leq f_k(\varphi)$ are global state
variables, and $T(w_{i,j},w_{i+1,j})$ is a formula encoding the
transition relation $T$.

An intuition behind this encoding is as follows.
The vector $w_{0,0}$ encodes the initial state $\iota$ and
for each symbolic $k$-path, numbered $1 \ldots f_k(\varphi)$, each pair 
of the consecutive vectors on this path encodes pairs of states that are 
in the transition relation $T$.
The formula $T(w,v)$ is typically a logical disjunction of the encodings 
of all the actions corresponding to the transitions of the model $M$.
This way, one symbolic $k$-path encodes all the (concrete) $k$-paths.

The next step of the algorithm consists in translating an \ECTLK\
formula $\varphi$ into a propositional formula. 
Let $w, v$ be global state variables. 
We make use of the following propositional formulas in the encoding:

\begin{itemize}

\item
$p(w)$ encodes a proposition $p$ of \ECTLK\ over $w$.

\item
$H(w, v)$ represents logical equivalence between global state
encodings $u$ and $v$  (i.e., encodes that $u$ and $v$ 
represent the same global states).

\item
$HK_i(w,v)$ represents logical equivalence between $i$-local
state encodings $u$ and $v$,
(i.e., encodes that $u$ and $v$ share $i$-local states).

\item
$L_{k,j}(l)$ encodes a backward loop connecting the $k$-th state
to the $l$-th state in the symbolic $k$-path $j$, 
for $0 \leq l \leq k$.
\end{itemize}
The translation of each \ECTLK\ formula is directly based on 
its bounded semantics.
The translation of $\varphi$ at the state $w_{m,n}$ into the
propositional formula ${[\varphi]}^{[m,n]}_{k}$ is as follows (we
give the translation of selected formulas only):

\begin{tabular}{lll}

$\!\!\!\!\!\!\!\!\!\!{[\E \X \alpha]}^{[m,n]}_{k}$ & := &
    $\bigvee_{i=1}^{f_k(\varphi)}\Big( H(w_{m,n},w_{0,i}) \;
    \land\; [\alpha]^{[1,i]}_{k} \Big)$,\\

$\!\!\!\!\!\!\!\!\!\!{[\E \G \alpha]}^{[m,n]}_{k}$&:=&
    $\bigvee_{i=1}^{f_k(\varphi)}\Big( H(w_{m,n},w_{0,i}) \;
    \land \; (\bigvee_{l=0}^{k} L_{k,i}(l))\; \land \; \bigwedge^{k}_{j = 0}
    [\alpha]^{[j,i]}_{k}\Big)$,\\

$\!\!\!\!\!\!\!\!\!\!{[\E (\alpha \U \beta)]}^{[m,n]}_{k}$& := &
    $\bigvee_{i=1}^{f_k(\varphi)} \Big( H(w_{m,n},w_{0,i}) \;
    \land \; \bigvee^{k}_{j=0} \big([\beta]^{[j,i]}_{k} \; \land \;
    \bigwedge^{j-1}_{t=0} [\alpha ]^{[t,i]}_{k}\big)\Big),$\\

$\!\!\!\!\!\!\!\!\!\!{[\NK_l\alpha]}^{[m,n]}_{k}$& := &
    $\bigvee_{i=1}^{f_k(\varphi)} \Big( I_{ g^0}(w_{0,i}) \;
   \land \;  \bigvee^{k}_{j=0} \big([\alpha]^{[j,i]}_{k} \; \land \;
   HK_l(w_{m,n},w_{j,i})\big)\Big).$\\

\end{tabular}

\noindent
Intuitively, $[\E \G \alpha]^{[m,n]_{k}}$ is translated to all the $f_k(\varphi)$-symbolic $k$-paths ($\E \G \alpha$ is considered 
as a subformula of $\varphi$) that start at the states encoded 
by $w_{m,n}$, satisfy $\alpha$, and are loops.   
$[\NK_l \alpha]^{[m,n]}_{k}$ is translated to all the $f_k(\varphi)$-symbolic
$k$-paths such that each symbolic $k$-path starts at the initial state $\iota$, 
one of its states satisfies $\alpha$ and shares the $l$-th state with these 
encoded by $w_{m,n}$.    
Given the translations above \cite{wlp-lcmas04}, verification 
of $\varphi$ over $M_k$ reduces to checking the satisfiability of 
the propositional formula $[M^{\varphi, g^0}]_k \land [\varphi]_{M_k}$, 
where $[\varphi]_{M_k} = [\varphi]^{[0,0]}_{k}$.

\subsection{Unbounded Model Checking}
\label{umc}

\UMC\ was originally introduced for verification of \CTL\ \cite{McMillan02} 
as an alternative to \BMC\ and approaches based on BDDs.
Then, \UMC\ was extended to \CTLpK\ \cite{KLP-fi04} 
as well as to other more expressive logics.

We begin by extending the syntax and semantics of \CTLK\ to \CTLpK\
by adding past operators $\A\Y$ and $\A\H$.  The operators including
Since are omitted.  A backward path $\pi = (g_0,g_1,\ldots)$ is a
maximal sequence of global states such that $(g_{k+1},g_{k}) \in T$
for each $k \geq 0$ (if $\pi$ is finite, then $k$ needs to be
restricted accordingly).  Let $\overline{\Pi}(g)$ denote the set of
all the backward paths starting at $g \in G$.
 
\begin{itemize}
\item $(M,g) \sat \A\Y \phi$ iff for all $\pi \in \overline{\Pi}(g)$ we have
      $(M,\pi(1)) \sat \phi$,
\item $(M,g) \sat \A\H \phi$ iff for all $\pi \in \overline{\Pi}(g)$ and for all
      $k \geq 0$ we have $(M,\pi(k)) \sat \phi$.
\end{itemize}
Unlike \BMC, \UMC\ is capable of handling the whole language of the logic. 
Our aim is to translate \CE\ formulas into propositional formulas
in conjunctive normal form, accepted as an input by SAT-solvers.
 
Specifically, for a given \CE\ formula $\varphi$, a corresponding 
propositional formula $[\varphi](w)$ is computed, where $w$ is 
a global state variable (i.e., a vector of propositional 
variables for representing global states) encoding these
states of the model where $\varphi$ holds.
The translation is not operating directly on temporal-epistemic formulas.
Instead, to calculate propositional formulas either the \QBF\ 
or the fix-point characterisation of \CE\ formulas 
(see Section \ref{bdd}) is used. 
More specifically, three basic algorithms are exploited.
The first one, implemented by the procedure {\it forall} \cite{McMillan02},
is used for translating formulas $\O\alpha$ such that  
$\O \in \{\A\X$, $\A\Y$, $K_{i}$, $D_{\Gamma}$, $E_{\Gamma}\}$.  
This procedure eliminates the universal quantifiers from 
a \QBF\ formula characterising a \CE\ formula, and returns 
the result in a conjunctive normal form.
The second algorithm, implemented by the procedure {\it gfp}$_O$
is applied to formulas $\O\alpha$ such that 
$\O \in \{\A\G,\A\H,C_{\Gamma}\}$.
This procedure computes the greatest fix-point, in the standard way,
using Boolean representations of sets rather than sets themselves.
For formulas of the form $\A(\alpha\U\beta)$ the third procedure, 
called {\it lfp}$_{AU}$, computing the least fix-point (in a similar way), is used.
In so doing, given a formula $\varphi$ a propositional formula
$[\varphi](w)$ is obtained such that $\varphi$ is valid in the model $M$ iff 
the propositional formula $[\varphi](w) \wedge I_{g^0}(w)$ is satisfiable. 

The reader is referred to \cite{KLP03-techreport} for more details,
especially on computing fix-points over propositional representations
of sets.
In the following section we show how to represent \CE\ formulas in
\QBF\ and then translate them to propositional formulas in \CNF.

\subsubsection{From a fragment of \QBF\ to \CNF}
\label{cnf-qbf}

\emph{Quantified Boolean Formulas} (\QBF) are an extension of propositional 
logic by means of quantifiers ranging over propositions.
The BNF syntax of a \QBF\ formula is given by: 

$$
\alpha::= p \mid \lnot \alpha \mid \alpha \land \alpha \mid 
\exists p.\alpha \mid \forall p.\alpha.
$$ 
The semantics of the quantifiers is defined as follows:

\begin{itemize}
\item[$\bullet$] $\exists p.\alpha$ iff $\alpha(p \leftarrow \true) \vee
                  \alpha(p \leftarrow \false)$,
\item[$\bullet$] $\forall p.\alpha$ iff $\alpha(p \leftarrow \true) \wedge
                  \alpha(p \leftarrow \false)$,
\end{itemize}
where $\alpha \in {\rm QBF}$, $p\in\PV$ and $\alpha(p \leftarrow q)$
denotes substitution with the variable $q$ of every occurrence of the
variable $p$ in formula $\alpha$. 
For example, the formula $[\A\X\alpha](w)$ is equivalent to the 
formula $\forall v. (T(w,v) \Rightarrow [\alpha](v))$ in \QBF.  
Similar equivalences are obtained for the formulas $\A\Y\alpha$, 
$K_i\alpha$, $D_\Gamma\alpha$, and  $E_\Gamma\alpha$ by 
replacing $T(w,v)$ with suitable encodings of the relations 
$T^{-1}$, $\sim_i$, $\sim^D_\Gamma$, and $\sim^E_\Gamma$.

\newcommand{\mi}[1]{\mathit{#1}}

For defining a translation from a fragment of \QBF\
(resulting from the translation of \CE)
to propositional logic, one needs to know how to compute a \CNF\ 
formula which is equivalent to a given propositional formula $\varphi$. 
While the standard algorithm $\tCNF$ \cite{McMillan02,pp-mono}, 
which transforms a propositional formula to one in \CNF, preserving 
satisfiability only, is of linear complexity, a translation to an equivalent 
formula is NP-complete.
For such a translation, one can use the algorithm {\em equCNF} - a version 
of  the algorithm $\tCNF$, known as a {\em cube reduction}. 
We refer the reader to \cite{cck-satbased-image,gga-umc04}, where 
alternative solutions can be found. 
The algorithm {\em equCNF} is a slight modification of the DPLL algorithm 
checking satisfiability of a \CNF\ formula (see \cite{pp-mono}), but it can be presented in a general way, abstracting away from its specific realisation. 

Assume that $\varphi$ is an input formula.
Initially, the algorithm {\em equCNF} builds a satisfying assignment for 
the formula ${\tCNF}(\varphi)\wedge \neg l_{\varphi}$ 
($l_{\varphi}$ is a literal used in ${\tCNF}(\varphi)$), 
i.e., the assignment which falsifies $\varphi$. 
If one is found, instead of terminating, the algorithm
constructs a new clause that is in conflict with the current
assignment (i.e., it rules out the satisfying assignment). 
Each time a satisfying assignment is obtained, a blocking clause 
is generated by a procedure {\tt blocking\_clause} 
and added to the working set of clauses. 
This clause rules out a set of cases where $\varphi$ is false. 
Thus, on termination, when there is no satisfying assignment for
the current set of clauses, the conjunction of the blocking 
clauses generated precisely characterises $\varphi$.

A blocking clause could in principle be generated using the conflict-based 
learning procedure.
If we require a blocking clause to contain only input variables,
i.e., literals used in $\varphi$, then one could either use an (alternative)
implication graph \cite{McMillan02} in which all the roots are input 
literals or a method introduced by Szreter \cite{ms-cfv05,ms-atva05}, which 
consists in searching a directed acyclic graph representing the formula.

Our aim is to compute a propositional formula equivalent to 
a given \QBF\ formula $\forall p_1\ldots\forall p_n.\varphi$.
The algorithm constructs a formula $\psi$ equivalent to 
$\varphi$ and eliminates from $\psi$ the quantified variables on-the-fly,
which is correct as $\psi$ is in \CNF.
The algorithm differs from {\em equCNF} in one step only, where 
the procedure {\tt blocking\_clause} generates a blocking clause 
and deprives it of the quantified propositional variables.
On termination, the resulting formula is a conjunction of the blocking 
clauses without the quantified propositions and precisely 
characterises $\forall p_1\ldots\forall p_n.\varphi$ 
(see \cite{KLP03-techreport,pp-mono} for more details).

\subsection{\verics}

\verics\ \cite{verics,Verics-csp04} is a verification tool 
for real-time systems (RTS) and \mas s (MAS). 
It offers three complementary methods of model checking: 
SAT-based Bounded Model Checking (\BMC), 
SAT-based Unbounded Model Checking (\UMC), 
and an on-the-fly verification while constructing abstract models of systems. 
The theoretical background for its implementation has been 
presented elsewhere \cite{pp-mono,pwz02}.

A network of communicating (timed) automata (together with a valuation function) 
is the basic \verics's formalism for modelling a system to be verified.  
Timed automata are used to specify RTS, whereas timed 
or untimed automata are applied to model MAS.
\verics\ translates a network of automata and a temporal-epistemic 
formula into a propositional formula in \CNF\ and invokes 
a \SAT-solver in order to check for its satisfiability. 

Currently, \verics\ implements \BMC\ for \ECTLKD\ 
(\ECTLK\ extended with deontic operators) 
and \TECTLK\ (see Section \ref{tctlk}),
and \UMC\ for \CTLpK.
\verics\ has been implemented in C++; its internal
functionalities are available via an interface written in 
Java \cite{verics-http}.  

\section{Extensions to real-time epistemic logic}
\label{tctlk}

In this section we briefly discuss some extensions to real-time to the
\ECTLK\ framework analysed so far.  The timed temporal-epistemic logic
\TECTLK\ \cite{lpw-AIJ07} was introduced to deal with situation where
time is best assumed to be dense and hence modelled by real numbers.
The underlying semantics uses networks of \emph{timed automata}
\cite{alur-dill94} to specify the behaviour of the agents.  These
automata extend standard finite state automata by a set of clocks
$\Cl$ (to measure the flow of time) and time constrains built over
$\Cl$ that can be used for defining guards on the transitions as well
invariants on their locations.  When moving from a state to another, a
timed automaton can either execute action transitions constrained by
guards and invariants, or time transitions constrained by invariants
only.  Crucial for automated verification of timed automata is the
definition of an equivalence relation $\equiv\; \subseteq \RR^{|\Cl|}
\times \RR^{|\Cl|}$ on clocks valuations, which identifies two
valuations $v$ and $v'$ in which either all the clocks exceed some
value $c_{max}$,\footnote{This constant is computed from a timed
  automaton and a formula to be verified.} or two clocks $x$ and $y$
with the same integer part in $v$ and $v'$ and either their fractional
parts are equal to $0$, or are ordered in the same way, i.e.,
$\mathit{fractional}(v(x)) \leq \mathit{fractional}(v(y))$ iff $\mathit{fractional}(v'(x)) \leq
\mathit{fractional}(v'(y))$.  The equivalence classes of $\equiv$ are called
{\em zones}.  Since $\equiv$ is of finite index, there is only
finitely many zones for each timed automaton.

In addition to the standard epistemic operators, the language 
of \TECTLK\ contains the temporal operators $\E\G$ and $\E\U$  
combined with time intervals $I$ on reals in order to specify 
when precisely formulas are supposed to hold. 
Note that \TECTLK\ does not include the next step operator $\E\X$ as this operator 
is meaningless on dense time models. 
The formal syntax of \TECTLK\ in BNF is as follows:
$$
\phi,\psi::= 
p\in \PV\ \mid \lnot p \mid \psi \land \phi \mid \psi \lor \phi\mid 
\NK_i \phi \mid \NE_{\Gamma} \phi \mid \ND_{\Gamma} \phi \mid \NC_{\Gamma} \phi
\mid \E\G_I \phi \mid \E(\phi \U_I \psi) 
$$
A (real time interpreted) model for \TECTLK\ over a timed automaton is
defined as a tuple $M=(Q,s^0,T,\sim_1,\ldots,\sim_n,V)$, where $Q$ is
the subset of $G \times \RR^{|\Cl|}$ such that $G$ is the set of
locations of the timed automaton, all the states in $Q$ are reachable
from $s^0 = (g^0,v^0)$ with $g^0$ being the initial location of the
timed automaton and $v^0$ the valuation in which all the clocks are
equal to $0$; $T$ is defined by the action and timed transitions of
the timed automaton, $\sim_i\; \subseteq Q \times Q$ is an epistemic
relation for agent $i$ defined by $(g,v) \sim_i (g',v)$ iff $g \sim_i
g'$ and $v \equiv v'$; and $V: Q \times \PV\ \to \{true, false\}$ is a
valuation function for $\PV$.  Intuitively, in the above model two
states are in the epistemic relation for agent $i$ if their locations
are in this relation according to the standard definition in Section
\ref{ss} and their clocks valuations belong to the same zone.

In what follows, we give the semantics of $\E(\phi\U_I\psi)$
and $\E\G_I\phi$ of \TECTLK\ and discuss how \BMC\ is applied to this logic.  
Differently from the paths of temporal-epistemic models,
the paths in real time models consist of action transitions interleaved 
with timed transitions.  
The time distance to a state $s$ from the initial one at a given path can 
be computed by adding the times of all the timed transitions that has 
occurred up to this state.
Following this intuition the semantics is formulated as follows: 
\begin{itemize}
\item
$(M,s) \models \E (\phi\U_I\psi)$ iff
there is a path in $M$ starting at $s$ which contains a state where
$\psi$ holds, reached from $s$ within the time distance of $I$, and
$\phi$ holds at all the earlier states,
\item
$(M,s) \models \E\G_I\phi$ iff there is a path in $M$ starting at $s$ such that 
$\phi$ holds at all the states within the time distance of $I$.
\end{itemize}
The idea of \BMC\ for $(M,s^0) \models \varphi$, where $\varphi$ is
\TECTLK\ formula, is based on two translations and on the application
of \BMC\ for \ECTLK.  An infinite real time model $M$ is translated to
a finite epistemic model $M_d$ and each formula $\varphi$ of \TECTLK\
is translated to the formula $cr(\varphi)$ of the logic \ECTLKy, 
which is a slight modification of \ECTLK.  
The above two translations guarantee that
$(M,s^0) \models \varphi$ iff $(M_d,s^0) \models cr(\varphi)$.

Assume we are given a timed automaton $A$ and a \TECTLK\ formula $\varphi$.  
We begin by translating the real time model $M$ (for $A$) to $M_d$.  
First, the automaton $A$ is extended with one special clock $y$, an action $a_y$, 
and the set of transitions $E_y$ going from each location to itself and resetting 
the clock $y$.  
These transitions are used to start the paths over which sub-formulas 
of $\varphi$ are checked.  
Then, the finite model $M_d$ for the extended timed automaton is built.  
The model $M_d = (Q_d,q^0,T_d,\sim^d_1,\ldots,\sim^d_n,\V_d)$, where $Q_d$ is a
suitably selected (via discretization) finite subset of $Q$, the
relations $T_d, \sim^d_i$ are suitably defined restrictions of the
corresponding relations in $M$, and $\V_d = \V|Q_d$.
  
The above translation $cr$ of the temporal modalities is non-trivial only. 
Applying $cr$ to $\E(\alpha\U_I\beta)$ we get the formula 
$\E\X_y\E(cr(\alpha)\U cr((\beta) \land p))$, where the operator 
$\E\X_y$ is interpreted over the transitions corresponding to the action 
$a_y$, and $p$ is a propositional formula characterising zones.
A similar translation applies to $\E\G_I\alpha$.

After the above two translations have been defined, the model checking
of a \TECTLK\ formula $\varphi$ over $M$ is reduced to model checking
of $cr(\varphi)$ over $M_d$, for which \BMC\ can be used as presented
in Section \ref{bmc}.

\subsection{Example}
\label{example-rcs}

To exemplify the expressive power of \TECTLK\ we specify a correctness
property for an extension of the {\em Railroad Crossing System} (RCS)
\cite{KangLee}, a well-known example in the literature of real time
verification.  
Below, we summarise the description from \cite{lpw-AIJ07}.

The system consists of three agents: Train, Gate, and Controller
running in parallel and synchronising through the events: 
{\it approach, exit, lower} and {\it raise}.  
When a train approaches the crossing, Train sends the signal 
{{\bf approach}} to Controller and enters the crossing between $300$ and $500$ 
milliseconds (ms) from this event.  
When Train leaves the crossing, it sends the signal {\bf exit} 
to Controller. Controller sends the signal {\bf lower} to Gate exactly
$100$ms after the signal {\bf approach} is received, 
and sends the signal {\bf raise} signal within $100$ms after {\bf exit}. 
Gate performs the transition {\bf down} within $100$ms of receiving 
the request {\bf lower}, and responds to {\bf raise} by moving 
{\bf up} between $100$ms and $200$ms.

Consider the following correctness property: 
there exists a behaviour of RCS such that agent Train considers 
possible a situation in which it sends the signal {\bf approach} 
but agent Gate does not send the signal {\bf down} within $50$ ms.
This property can be formalised by the following \TECTLK\ formula: 
$$\varphi
=\E\F_{[0, \infty]}\NK_{Train}({\bf approach} \land \E\F_{[0,50]}(\neg {\bf down})).
$$
By using \BMC\ techniques we can verify the above property for RCS.

\section{Related Work}
\label{relatedwork}

The approaches above have been extended in several directions and
other articles have appeared pursuing related lines.  It was mentioned
in Section~3 that van der Meyden and colleagues were the first to
propose concretely how \obdds\ could be used to model check
temporal-epistemic logic as well as to study the complexity of the
model checking problem in specific cases \cite{meydenshilov99}
(\cite{LomuscioRaimondi06b} has further results on this).  As
discussed above the main difference of their approach to the one
presented here is the different semantics employed and the particular
optimisation techniques used on it.  We refer to \cite{mck04,mck-tool}
for more details.  We are not aware of other symbolic efforts other
than the one presented above as far as SAT-based techniques (\BMC,
\UMC) are concerned.  However, different techniques for
temporal-epistemic logic have been put forward in the past.  

In \cite{HoekWooldridge02a} van der Hoek and Wooldridge suggested
reduction of temporal-epistemic logic to temporal logic only by using
local propositions fully describing agents' local states. The approach
consists in manually finding appropriate propositions describing
appropriate states.  An example of the technique is described on an
example in \cite{HoekWooldridge03b} where the ATL model checker MOCHA
\cite{MOCHAtechrep2000} is used (see also below).  Lastly,
temporal-epistemic logic on discrete time was recently recast as a
special case of ARCTL \cite{PecheurRaimondi06}.  An extension of NuSMV
was introduced to implement ARCTL \cite{PecheurRaimondi06} thereby
enabling the verification of CTLK directly on NuSMV via an ad-hoc
translation as discussed in \cite{Lomuscio+07}.

Model checking has also been investigated for certain extensions of
the temporal-epistemic logics discussed here.  In
\cite{RaimondiLomuscio04c} an OBDD-based approach to the verification
of deontic interpreted systems \cite{LomuscioSergot03a} is presented
and in \cite{wlp-lcmas04} the \BMC\ case was analysed.  Deontic
interpreted systems are a formalism enabling the representation and
the distinction of correct versus incorrect states of agents.  In this
framework local states are partitioned into correct and incorrect
local states and a modality $O_i$ introduced for every agent
evaluating formulas only at the correct states thereby representing
concepts such as ``all the correct states for agent $i$''.  For
instance, one could analyse a variant of the dining cryptographers
scenario where some cryptographers are intruders saying the opposite
of what they should \cite{Kacprzak+06}.  Extensions to epistemic logic
to include explicit knowledge have also been discussed and implemented
\cite{LomuscioWozna06a,Lomuscio+07b}.  Both \verics\ and MCMAS support
these formalisms.

In other developments model checking of epistemic logic in an ATL
\cite{atl02} setting has also been pursued.  ATL can be shown to
extend \CTL\ (at some computational cost) by adding strategies in the
semantics and explicit representation of the notion of enforcement in
the syntax. Even if strategies and knowledge can interact in rather
subtle ways \cite{Jamroga-Hoek}, progress has been made both in the
definition of ATL extensions including knowledge and other modalities
and in their verification.  We refer to \cite{Wooldridge+07} for an
up-to-date survey and references.  The approach taken there uses MOCHA
\cite{MOCHAtechrep2000} and the local propositions construction
referenced above.  MCMAS described earlier in this survey also supports
ATL natively in the different knowledge semantics proposed.  We do not
discuss the syntax here and refer to above mentioned references for
more details.

Elsewhere epistemic-like concepts have been used in a broader context
to reason about multi-agent systems modelled by other attitudes (such
as norms, beliefs, desires, or intentions). Normally these properties
are treated simply as propositions in a temporal language and not as
prima-specie citizens like the epistemic modalities above,
consequently the approaches are rather different and not discussed here.

\section{Conclusions}
\label{con}

It has long been argued that epistemic logic provides an intuitive
formalism in a variety of key areas in computer science.  In this
article we have surveyed some of the recent contributions to solving
the model checking problem for temporal-epistemic logic in a branching
time setting under a discrete and a continuous model of time.
The conclusion we can draw from the above is that model checking
temporal-epistemic logic is very often no harder than plain temporal
logic; however most procedures and particular algorithms need to be
extended to accommodate this need. 
Now that model checking algorithms and tools have been made available
it will be interesting to see the extent to which temporal-epistemic
logic can be used in real-life scenarios.

\paragraph{Note.} The techniques described in Sections 3--5 were
joint work of the authors with M. Kacprzak, F. Raimondi, and
B. Wo\'{z}na.

\end{document}